\documentclass[aps,prd,twocolumn,floatfix,preprintnumbers,altaffilletter]{revtex4-1}

\usepackage{graphicx,url,amssymb,amsmath,longtable,rotating,color,units,wasysym,subfigure,epsfig,multirow,epstopdf}
\usepackage[colorlinks,urlcolor=blue,citecolor=blue,linkcolor=blue]{hyperref}

\definecolor{darkgreen}{rgb}{0.01, 0.75, 0.24}

\begin{document}

\title{Suspending test masses in terrestrial millihertz gravitational-wave detectors:\\ a case study with a magnetic assisted torsion pendulum}

\author{Eric~Thrane}
\email{eric.thrane@monash.edu}
\affiliation{School of Physics and Astronomy, Monash University, Clayton, Victoria 3800, Australia}
\affiliation{OzGrav: The ARC Centre of Excellence for Gravitational-wave Astronomy, Hawthorn, Victoria 3122, Australia}

\author{R.~P.~Anderson}
\affiliation{School of Physics and Astronomy, Monash University, Clayton, Victoria 3800, Australia}

\author{Yuri Levin}
\affiliation{School of Physics and Astronomy, Monash University, Clayton, Victoria 3800, Australia}
\affiliation{OzGrav: The ARC Centre of Excellence for Gravitational-wave Astronomy, Hawthorn, Victoria 3122, Australia}

\author{L.~D.~Turner}
\affiliation{School of Physics and Astronomy, Monash University, Clayton, Victoria 3800, Australia}

\date{\today}

\begin{abstract}
  Current terrestrial gravitational-wave detectors operate at frequencies above $\unit[10]{Hz}$.
  There is strong astrophysical motivation to construct low-frequency gravitational-wave detectors capable of observing $\unit[10]{mHz}$--$\unit[10]{Hz}$ signals.
  While space-based detectors provide one means of achieving this end, one may also consider terretrial detectors.
  However, there are numerous technological challenges.
  In particular, it is difficult to isolate test masses so that they are both seismically isolated and freely falling under the influence of gravity at millihertz frequencies.
  We investigate the challenges of low-frequency suspension in a hypothetical terrestrial detector.
  As a case study, we consider a Magnetically Assisted Gravitational-wave Pendulum Intorsion (MAGPI) suspension design.
  We construct a noise budget to estimate some of the required specifications.
  In doing so, we identify what are likely to be a number of generic limiting noise sources for terrestrial millihertz gravitational-wave suspension systems (as well as some peculiar to the MAGPI design).
  We highlight significant experimental challenges in order to argue that the development of millihertz suspensions will be a daunting task.
  Any system that relies on magnets faces even greater challenges.
  Entirely mechanical designs such as Z\"ollner pendulums may provide the best path forward.
\end{abstract}

\maketitle

{\em Introduction.}
Second-generation gravitational-wave detectors~\cite{aligo2,virgo,kagra} observe at frequencies above $f \approx \unit[10]{Hz}$.
This is because the test masses are suspended with linear pendulums with resonant frequencies $\approx \unit[1]{Hz}$, which isolate the test masses from seismic noise.
They also ensure that the test masses behave as though they are freely falling, which allows them to move under the influence of gravitational waves.
However, both seismic suppression and coupling to gravitational waves are only effective above the resonant frequency.
Below the resonant frequency, seismic noise is not attenuated and the coupling to gravitational waves falls rapidly.

There is strong astrophysical motivation to design gravitational-wave detectors that can operate at lower frequencies~\cite{mango}:
\begin{enumerate}
\item Gain sensitivity to intermediate-mass black-hole mergers with total mass $\gtrsim1600M_\odot$.
  \item Improve sensitivity to stochastic backgrounds. 
    Since the stochastic search signal-to-noise ratio scales like $f^{-3}$, it is possible to achieve a dramatic improvement in sensitivity by expanding the observation band to include lower frequencies~\cite{allenromano}.
  \item Gain sensitivity to the $\approx$80\% of pulsars that emit gravitational waves with $f < \unit[10]{Hz}$---out of the reach of audio-band detectors~\cite{atnf}.
  \item Improve low-latency follow-up alerts by gaining an advanced inspiral signal prior to merger~\cite{early_warning}.
\end{enumerate}

Despite this strong motivation, it is not easy to modify current detectors that are based on linear pendulums to probe frequencies below $\unit[10]{Hz}$.
This is because the resonant period of a linear pendulum scales with the square root of its length, and extremely long pendulums ($\text{length}\gg\unit[1]{m}$) are impractical.

The most plausible solution at present is to move the detector to space.
The recent success of LISA Pathfinder provides encouragement to continue on this path~\cite{lisa_pathfinder}.
Nonetheless, there are good reasons for thinking about terrestrial millihertz detectors as well.
First, it may be possible to cover an interesting observing band $\unit[0.1-10]{Hz}$, which sits in between the observing bands of LISA and LIGO.
Second, the difficulties of creating (and repairing) facilities in space should lead us to consider everything that can be done on Earth.

The MANGO proposal is a design sketch for a terrestrial millihertz detector~\cite{mango}.
The target strain sensitivity of MANGO is $\unit[10^{-20}]{Hz^{-1/2}}$ at $\unit[100]{mHz}$.
This target is extremely ambitious.
Numerous technical challenges must be overcome in order to come close.
In this article, we investigate just one component in the design of a terrestrial millihertz detector such as MANGO: the final stage of the suspension system depicted in Fig.~7 of~\cite{mango}.
Working within the MANGO framework, and making optimistic assumptions, we explore a possible solution to the problem of suspending test masses at low frequencies, which is just one of the many difficult challenges facing millihertz detectors.

The attenuation of seismic noise, and the coupling of gravitational waves to a suspended test mass, are both described by transfer functions.
For a linear pendulum, the seismic transfer function relating displacement noise $x_t$ (at the top suspension point from which the pendulum is hung) to displacement of the test mass at the bottom $x_b$ is given by
\begin{equation}
  T_s(f) = \frac{x_b}{x_t} = \frac{f_0^2}{f_0^2 - f^2} ,
\end{equation}
where $f$ is the measured gravitational-wave frequency and $f_0$ is the resonant frequency of the linear pendulum.
For the sake of simplicity, we ignore the imaginary dissipative term, which serves to broaden the resonant peak at $f_0$.

\begin{figure}[hbtp!]
 \includegraphics[height=2.5in]{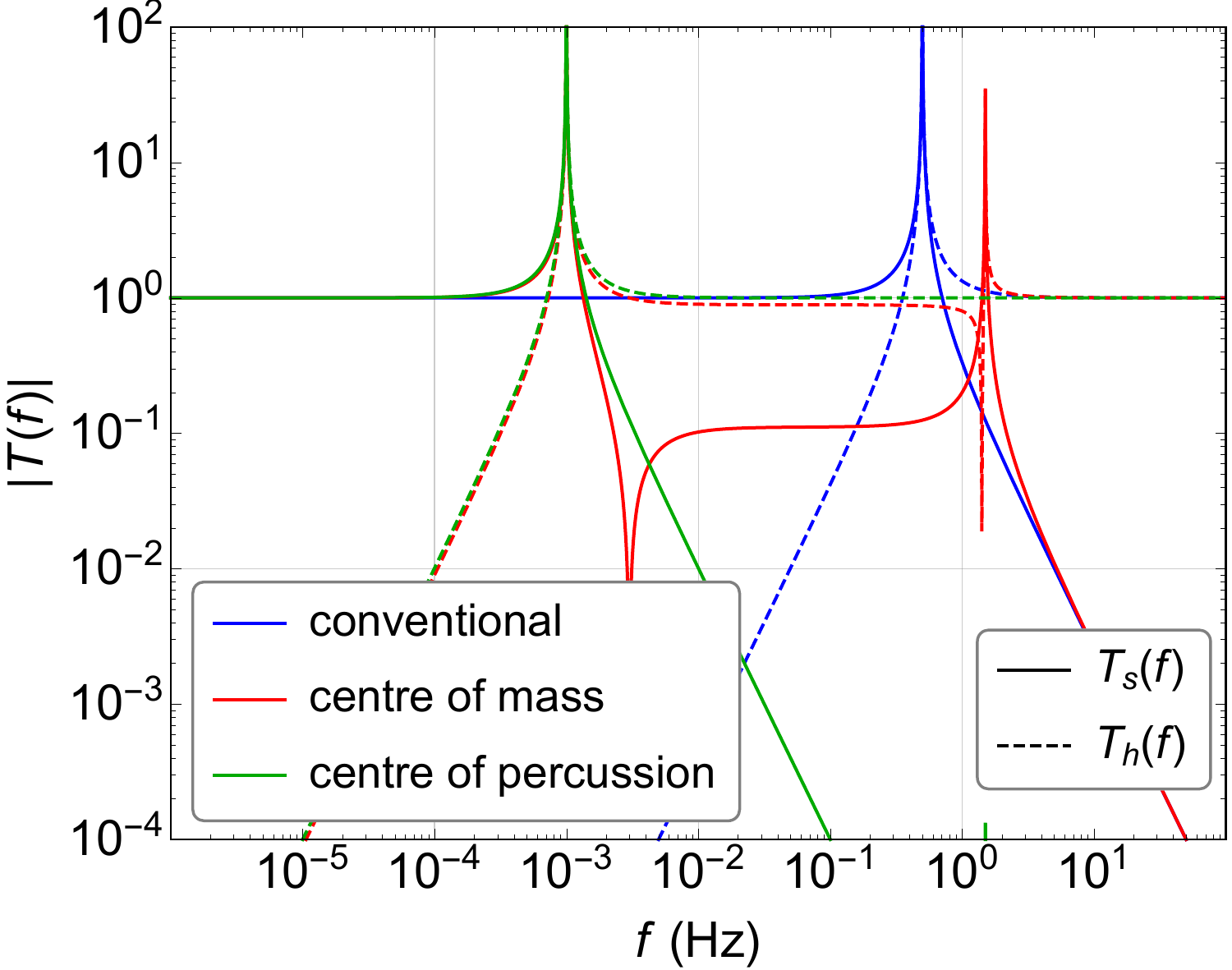}
  \caption{
    Transfer functions for seismic noise (solid) and gravitational waves (dashed).
    Blue is for a linear pendulum with resonant frequency $f_0=\unit[0.5]{Hz}$, red is for the center of mass in a magnetically assisted (MAGPI) pendulum $f_r=\unit[1]{mHz}$, and green is for the MAGPI center of percussion, offset from the center of mass.
  }
  \label{fig:plot_tf}
\end{figure}

The acceleration from a gravitational wave on an interferometer test mass can be written as $a_h=-4\pi^2f^2hL$ where $h$ is the gravitational-wave strain and $L$ is the length of the interferometer arms.
For a linear pendulum, the gravitational-wave transfer function relating $h$ to the measured strain $x_b/L$ is given by:
\begin{equation}
  T_h(f) = \frac{x_b}{hL} = \frac{f^2}{f^2-f_0^2} .
\end{equation}
In Fig.~\ref{fig:plot_tf} we plot $T_s(f)$ (solid) and $T_h(f)$ (dashed) for a (typical) $\unit[1]{m}$ linear pendulum with $f_0=\unit[0.5]{Hz}$ in blue.
We see that below $f_0$, the test mass is not isolated from seismic noise, and the coupling to gravitational waves falls like $f^{2}$.

Torsion pendulums are widely used in precision measurement because they can be made with very low resonant frequencies $f_0\lesssim\unit[1]{mHz}$ and with very high quality factors $Q\approx10^4$~\cite{hagedorn,adelberger}.
Ando \emph{et al.} have previously noted that a pair of torsion pendulums can be coupled to tidal deformations from gravitational waves, thereby exploiting the naturally low resonance frequency of torsion pendulums~\cite{toba}.
However, the sensitivity of the detector is limited by the $\unit[10]{m}$ length of the torsion bars.
See also~\cite{Mamakoukas} for a discussion of maglev-based suspension.

{\em A magnetically assisted torsion pendulum.}
In this paper we consider---as a case study---a magnetically assisted torsion pendulum for use in a LIGO-like interferometer.
We refer to this scheme as Magnetically Assisted Gravitational-wave Pendulum Intorsion (MAGPI).
We do not argue that this design is necessarily the best method of suspension in a millihertz detector.
Rather, the point of our study is to consider {\em one} concrete design in order to estimate a noise budget.
Put differently, the paper is intended as a conversation-starter, not a mature design proposal.
Much of the discussion here (although not all) is likely to apply to other designs as well.
In the conclusions, we discuss alternative designs, some of which avoid the use of magnets.

\begin{figure}[hbtp!]
  \includegraphics[trim={1cm 2cm 0 0},clip,width=\columnwidth]{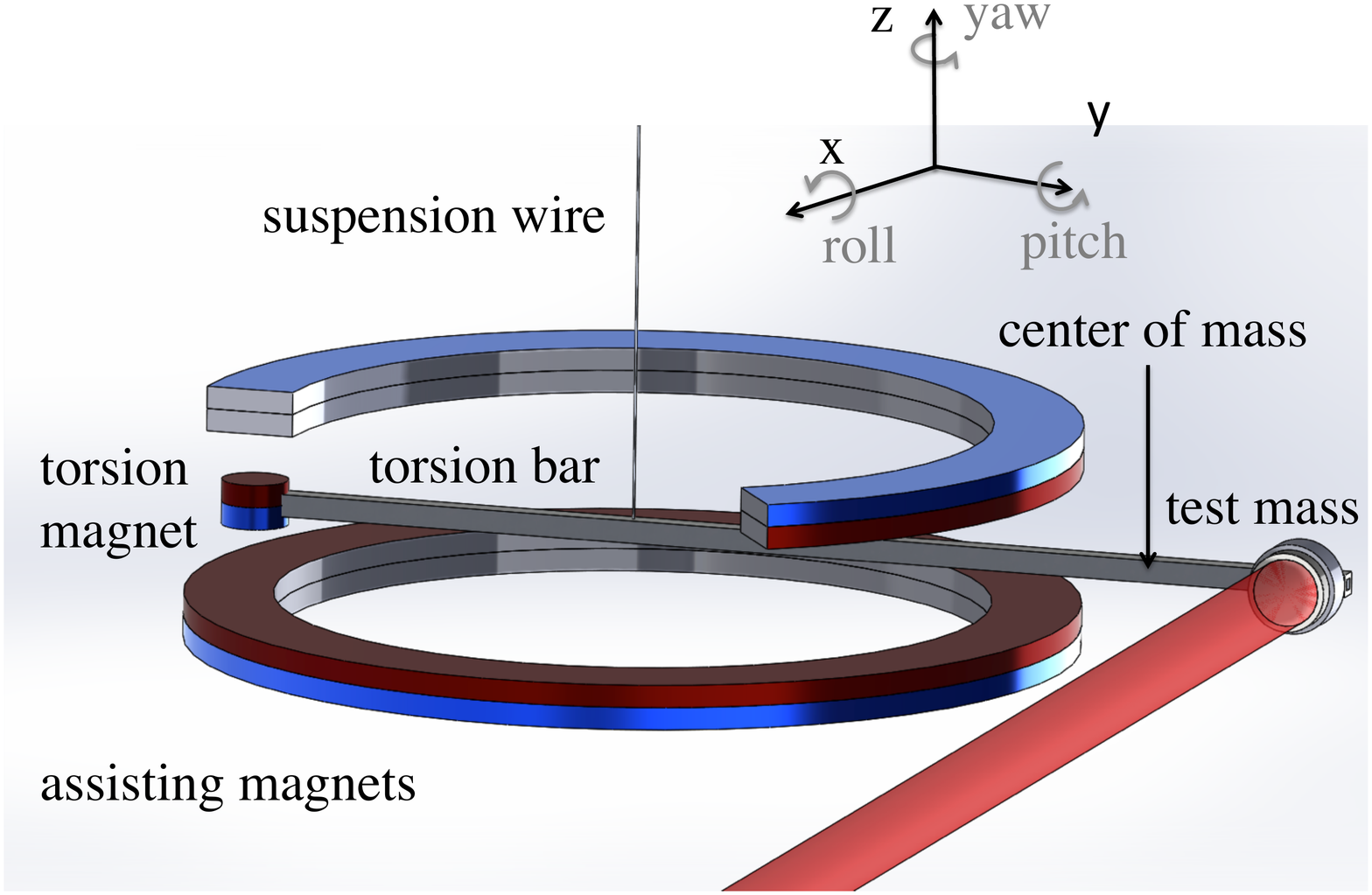}
  \caption{
    Schematic of the Magnetically Assisted Gravitational-wave Pendulum Intorsion (MAGPI); not to scale, with a sector of the upper assisting magnet made invisible.
    The assisting magnets exert a force on the torsion bar magnet, which holds up the bar.
    Without this torque, the bar would hang vertically.
    Gravitational-wave strain, acting on the center of mass, induces motion in the yaw degree of freedom.
  }
  \label{fig:magpi}
\end{figure}

A schematic of the MAGPI design is provided in Fig.~\ref{fig:magpi}.
We employ an asymmetric, unbalanced torsion bar with a test mass on one side and a much less massive permanent magnet on the other side.
We refer to this magnet as the ``torsion magnet'' to distinguish it from the ``assisting magnets,'' that exert a vertical magnetic force $\vec{F}=\mu \vec\nabla B_z$ on the torsion bar to hold it horizontal~\footnote{In order to avoid a yaw instability, it may be necessary to introduce a keel (not pictured). The keel extends underneath the torsion bar. If the torsion bar yaws slightly, the keel exerts a restoring torque, which balances the magnetic torque of the upper assisting magnet, which could otherwise flip the bar.}.
Here, $\vec\mu=\mu\hat{z}$ is the magnetic moment of the torsion magnet and $\vec{B}$ is the magnetic field of the assisting magnets.
Identical MAGPIs are situated at the ends of each arm of a Michelson interferometer.
The torsion bar and the assisting magnets are both suspended from a pre-isolation system consisting of a suspension-point interferometer and an optically rigid body as in~\cite{mango}.
In Fig.~\ref{fig:magpi}, a mirror mounted on the test mass reflects light in the $\hat{x}$ direction, down the interferometer arm.

In this asymmetric configuration, the center of mass is offset from the lower suspension point in the $\hat{y}$ direction, and so a gravitational wave will couple to the torsional (yaw) degree of freedom by exerting a torque on the bar.
As a consequence, the test mass behaves as though it is suspended by a low-resonant-frequency torsion pendulum, greatly expanding the frequency range over which gravitational waves can be measured.
To illustrate this, we calculate the seismic transfer function $T_s'(f)$ and the gravitational-wave transfer function $T_h'(f)$ for the MAGPI design.
The prime distinguishes the MAGPI transfer functions from those of the conventional linear pendulums discussed above.

For illustrative purposes, we assume the following fiducial parameters.
The test mass is $m=\unit[100]{kg}$.
The mass of the bar and the torsion magnet are assumed to be small in comparison.
The distance between the lower suspension point and the center of mass is $r=\unit[50]{cm}$.
For reasons described below, the test mass is placed $\unit[56]{cm}$ away from the lower suspension point, $\unit[6]{cm}$ beyond the center of mass.
In order to minimize noise from eddy currents, the magnets are constructed from a high-remanence, ferrimagnetic insulator such as yttrium iron garnet (YIG).

Considering the total force on the center of mass yields the following equation of motion:
\begin{equation}\label{eq:force}
  -f^2 x_m = -f_0^2(x_m+r\phi - x_t) -f^2 L h,
\end{equation}
where $x_m$ is the displacement of the center of mass along $\hat{x}$, $f$ is frequency, $f_0=\unit[0.5]{Hz}$ is the resonant frequency of the linear pendulum degree of freedom, $x_t$ is the seismically-induced displacement at the upper suspension point along $\hat{x}$, and $\phi$ is the (right-hand) yaw angle about the $\hat{z}$ axis.
Considering the total torque about the center of mass yields another equation of motion:
\begin{equation}\label{eq:torque}
  -I f^2\phi = -rm f_0^2(x_m + r\phi - x_t) - I_z f_r^2\phi .
\end{equation}
Here, $f_r \approx \unit[1]{mHz}$ is the rotational resonant frequency of the torsion pendulum, $I$ is the yaw moment of inertia about the center of mass, and $I_z\approx I+mr^2$ is the moment of inertia about the lower suspension point.

Combining Eqs.~\ref{eq:force}-\ref{eq:torque}, we obtain the transfer functions:
\begin{equation}
  \begin{split}
    T_s'(f) \equiv \frac{x_m}{x_t} = & \frac{f_0^2 \left(\kappa f_r^2 - f^2\right)}{f^4 - f^2 \kappa \left(f_0^2+f_r^2\right)+\kappa f_0^2 f_r^2} \\
    T_h'(f) \equiv \frac{x_m}{hL} = & \frac{f^2 \left(f^2-\kappa \left(f_0^2+f_r^2\right)+f_0^2\right)}{f^4-\kappa f^2 \left(f_0^2+f_r^2\right)+\kappa f_0^2 f_r^2}
    \end{split}
\end{equation}
Here, we introduce $\kappa=mr^2/I+1$, which becomes larger as mass is more efficiently concentrated around the center of mass.
For our design, $\kappa\approx 9$.

In Fig.~\ref{fig:plot_tf}, we plot $T_s'(f)$ (solid) and $T_h'(f)$ (dashed) in red.
We see that the test mass behaves as though it is freely falling (along $\hat{x}$) above $f_r=\unit[1]{mHz}$.
However, the seismic noise between $(f_r,f_0)$ is only suppressed by a factor of $\approx 9$.
In order to further attenuate seismic noise, we consider a point some distance $\delta r$ beyond the center of mass known as the center of percussion:
\begin{equation}
  \delta r = \frac{r}{\kappa-1} \left(1 - \kappa\frac{f_r^2}{f_*^2}\right) .
\end{equation}
At the center of percussion, the seismic coupling is zero when $f=f_*$.
Physically, the test mass is at rest while the bar rotates back and forth about it.
We treat $f_*$ as a tunable parameter that varies with $\delta r$.
However, the best broadband suppression of seismic noise occurs when $f_* \gg f_r$, in which case $\delta r \rightarrow r / (\kappa-1)$.
For the design considered here, $\delta r = \unit[6]{cm}$.

At the center of percussion, the transfer functions become
\begin{equation}\label{eq:doubleprime}
  \begin{split}
     T_s''(f) = & \frac{\kappa f_0^2 f_r^2 \left(f_*^2-f^2\right)}{f_*^2 \left(f^4-\kappa f^2 \left(f_0^2+f_r^2\right)+\kappa f_0^2 f_r^2\right)} 
         \rightarrow -\frac{f_r^2}{f^2} \\
     T_h''(f) = & \frac{f^2(f_*^2(f^2-\kappa(f_0^2+f_r^2))+\kappa f_0^2 f_r^2)}{f_*^2 \left(f^4-\kappa f^2 \left(f_0^2+f_r^2\right)+\kappa f_0^2 f_r^2\right)} \rightarrow 1
  \end{split}
\end{equation}
The double-prime denotes that we are considering motion at the center of percussion and not the center of mass.
The arrows in Eq.~\ref{eq:doubleprime} show the behavior in the limit that $f_0,f_* \gg f > f_r$.
Thus, a test mass placed at the center of percussion behaves as though it is suspended from a linear pendulum with resonant frequency $f_r$.
In Fig.~\ref{fig:plot_tf}, we plot $T''_s(f)$ (solid) and $T''_h(f)$ (dashed) for the MAGPI center of percussion (green).
It is interesting to consider this behavior qualitatively.
At frequencies below the linear pendulum resonance, the lower suspension point behaves as though it is rigidly attached.
Forces acting on the center of mass cause the test mass to yaw around the lower suspension point.
However, the induced motion at the center of percussion is comparatively small.
Having demonstrated that a torsional pendulum can be in principle coupled to the translational degree of freedom measured by a LIGO-like interferometer, we now turn our attention to noise.

There are several possibly-limiting noise sources for low-frequency gravitational-wave detectors including thermal noise, Newtonian gravity gradient noise, and radiation pressure noise~\cite{mango}.
For the sake of comparison, we assume that our MAGPIs are installed in an $L=\unit[300]{m}$ Michelson interferometer as per the MANGO design described in~\cite{mango}.
We further assume that the MAGPIs operate with a resonant frequency of $f_r=\unit[1]{mHz}$ so that seismic noise can be suppressed by the previous stages in the MANGO design: a suspension-point interferometer from which is hung an optically rigid body~\cite{mango}.
(The assisting magnets are suspended in addition to the torsion bar.)
The optically rigid body eliminates most seismic noise through a servo, which also eliminates gravitational-wave induced motion in the isolation stage.
The final MAGPI stage is necessary so that the test mass couples to gravitational waves.
We ignore Newtonian gravity gradient noise likely common to any terrestrial detector; see~\cite{driggers} for a possible mitigation strategy.

\begin{figure}[hbtp!]
  \includegraphics[width=3.0in]{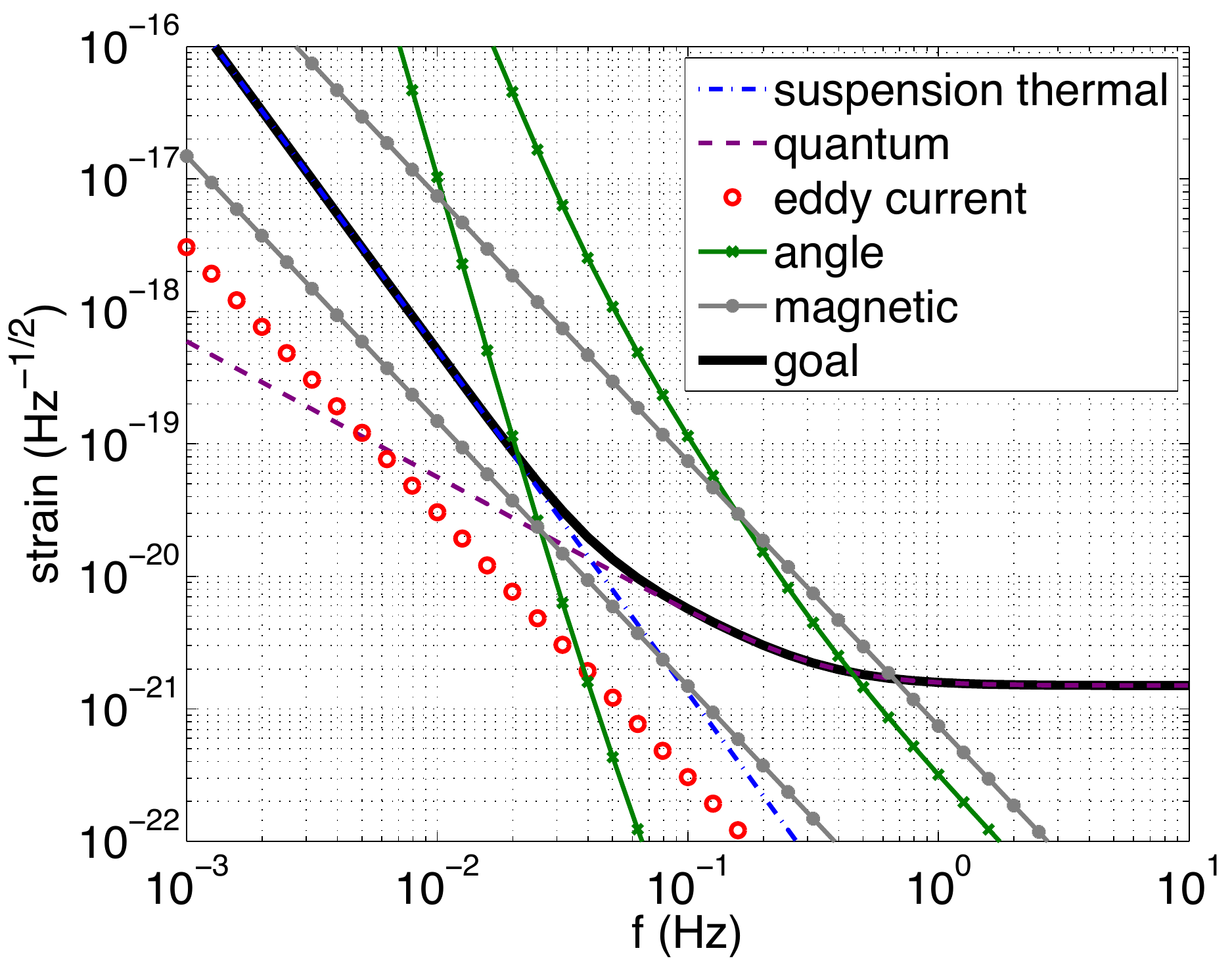}
  \caption{
    MAGPI noise budget.
    The dashed purple dash-dot blue curves are the assumed quantum noise and the suspension thermal noise taken from~\cite{mango}.
    The quadrature sum of these noises is the thick black line labeled ``goal'': we seek to avoid introducing noise above this level.
    The red $\circ$ show the expected noise from eddy currents.
    The green $\times$ show a range of possible noise from angular misalignment.
    The upper curve shows the noise if the MAGPI tilts along with the pre-isolation stage whereas the lower curve assumes that the MAGPI is further isolated by a mechanical filter with a resonant frequency of $\unit[1]{mHz}$; see the text for details.
    The gray $\bullet$ show a range of possible noise from ambient magnetic fields.
    The upper curve assumes a magnetic gradient noise of $\unit[0.1]{fT/m/Hz^{1/2}}$ achieved with a magnetic shield.
    The lower curve assumes that the magnetic noise is further reduced by a factor of $50$, e.g., with feed-forward; see the text for details.
  \label{fig:noisebudget}
  }
\end{figure}

Our noise budget is shown in Fig.~\ref{fig:noisebudget}.
In the MANGO design, the limiting noise below $\unit[30]{mHz}$ is due to suspension thermal noise (blue); above $\unit[30]{mHz}$ the limiting noise is quantum noise (purple).
The total noise is given by the black curve.
We use the MANGO estimates for these two noise sources as our starting point and investigate if the MAGPI design induces additional noise above this noise floor.
It is not obvious that the MANGO thermal noise budget applies to our torsion pendulum, but we will adopt it nonetheless in order to focus on additional non-thermal noise.
For additional details about the MANGO design, see Tab.~1 from~\cite{mango}.

{\em Dissipation by eddy currents} is a significant concern for any low-noise system utilizing strong magnetic fields.
Using the fluctuation-dissipation theorem, we investigate the scaling laws for eddy current noise following the formalism from~\cite{levin}.
The strain noise from eddy currents dissipated in the bar magnet can be expressed as
\begin{equation}\label{eq:eddy}
    \sigma_h \approx c\frac{1}{L} \sqrt{8 k_B T \sigma (\pi R^2 z)} \left(\frac{B}{m \omega^2}\right) .
\end{equation}
Here, $k_B$ is the Boltzmann constant, $T=\unit[293]{K}$ is temperature, $R\approx\unit[20]{mm}$ is the radius of the torsion magnet, $z\approx\unit[2]{cm}$ is the height of the torsion magnet, $\sigma$ is the conductivity of the magnets, $B$ is the magnetic field amplitude from the assisting magnets in the vicinity of the torsion magnet, and $c\lesssim1$ is a geometric constant.
We assume that the distance from the suspension point to the bar magnet is $\approx r+\delta r$.
In the calculations that follow, we set $c=1$.
A derivation of this equation is included in the appendix.

The ferrimagnetic material YIG has a large remanence but a low conductivity: $\sigma\approx\unit[10^{-11}]{S/m}$ at $T=\unit[293]{K}$, meaning that it can generate a relatively large magnetic field while suppressing eddy currents~\cite{sirdeshmukh}.
Given these assumptions, the eddy current noise in the bar magnet is $\sigma_h\lesssim\unit[10^{-22}]{Hz^{-1/2}}$ at $f=\unit[0.1]{Hz}$: well below other sources of noise.

Eddy current noise also arises from dissipation in the assisting magnets.
The volume of the assisting magnets is bigger than the volume of the torsion magnet, but the magnetic field is only large near the torsion magnet.
As an informed guess, therefore, we assume that the eddy current noise from each assisting magnet is about the same as from the torsion magnet, i.e., the total eddy current noise is $\sqrt{3}$ times the expression in Eq.~\ref{eq:eddy}.
Our rough approximation of the total eddy current noise is indicated with red circles in  Fig.~\ref{fig:noisebudget}.
We note that eddy current noise may be further reduced ($\gtrsim10\times$) through the use of laminations of the kind used in electrical transformers.

While eddy current dissipation in the ferrite magnets may not be a significant noise source, Eq.~\ref{eq:eddy} tells us that there are strict constraints on the proximity of conductors.
The conductivity of typical conductors such as copper is $\unit[6\times10^7]{S/m}$, eighteen orders of magnitude greater than YIG.
It is therefore difficult to envision an experimental apparatus in which conductors are used for actuation, e.g., electrostatic drives.
We therefore suspect that it would be necessary to carry out all actuation with other technologies, e.g., radiation pressure~\cite{pcal}.

{\em Angular alignment} noise, e.g., from tilt, has been previously identified as a limiting noise source for purely mechanical low-frequency suspension systems~\cite{garoi}.
For the MAGPI design, a non-zero pitch angle $\theta$ in any magnet (either the assisting magnets or the torsion magnet) will couple into the gravitational-wave channel via a yaw torque.
Alignment noise $\sigma_\theta(f)$ describes fluctuations in the pitch angle of the magnets.
The transfer function $T_\theta(f)=h/\theta$ is
\begin{equation}\label{eq:theta}
  T_\theta(f)
  \approx \left(\frac{r + \delta r}{L}\right) \left(\frac{f_\theta^2}{f_r^2 - f^2}\right)
\end{equation}
where $f_\theta=\sqrt{g/r}/2\pi\approx\unit[500]{mHz}$ and $g$ is the acceleration due to gravity.
The factor of $g$ appears because the magnetic torque is tuned to balance the gravitational torque.
The fact that $f_\theta\gg f_r$ implies that alignment noise will tend to be amplified through most of the observing band.
This places strict requirements on tilt noise.

The pre-isolation stage limits $\sigma_\theta(f)$ using active feedback.
By measuring the $\hat{x}$ displacement of the bottom and top of the optically rigid body, $\theta$ can (at best) be measured with uncertainty $\sigma_\theta=\sqrt{2}(\sigma_hL/\ell)$, where $\sigma_h$ is the total strain MANGO sensitivity and $\ell\approx\unit[1]{m}$ is the vertical distance between the bottom and top measuring points.
When $f=f_\theta$, $\sigma_\theta\approx\unit[8\times10^{-19}]{rad/Hz^{1/2}}$.

The transfer function describing how tilt at the pre-isolation stage $\theta_1$ couples to the tilt of the MAGPI magnets $\theta_2$ is $T_\theta'(f) = (f_\theta'/f)^2$.
Here, $f_\theta'=\sqrt{\mathfrak{r}mg/I_y}/2\pi$ depends on the distance between the center of mass and the suspension point $\mathfrak{r}$ and the pitch moment of inertia about the lower suspension point $I_y$.
Angular alignment noise can be reduced by making $\mathfrak{r}$ small, which reduces $f_\theta'$.
That is, there is a mechanical resonance, which can be used to attenuate pitch noise.
Tilt noise is converted into displacement noise at the suspension point.
Here, we suppose that $f_\theta'=\unit[1]{mHz}$, an optimistic but plausible value.
The green traces in Fig.~\ref{fig:noisebudget} show $\sigma_h^\theta$ at the pre-isolation stage (top) and after coupling to the MAGPI magnets (bottom).
We envision a similar kind of mechanical filter used to hang the assisting magnets, which, we presume would be suspended from the same optically rigid body as the torsion bar.

We anticipate alignment noise is a general problem for assisted pendulums because fluctuations in the assisting force will always couple at some level to the strain degree of freedom.
We hypothesize that, for any design, there is no ``free lunch'': the natural frequency of the assisting force will be close to the natural frequency of the unassisted pendulum, and that noise coupling through this mechanism will grow at low-$f$ like $\sigma_h\propto 1/f^2$.

{\em Ambient magnetic fields} from geophysical and anthropogenic sources create fluctuations at the level of $\sigma_B\approx\unit[1]{pT/Hz^{1/2}}$ in our frequency band~\cite{marfaing}.
Shielding can reduce magnetic noise to $\sigma_B\lesssim\unit[1]{fT/Hz^{1/2}}$~\cite{kornack}.
Thanks to the orientation of the torsion magnet, {\em uniform} magnetic fields do not efficiently couple to the yaw degree of freedom because the torque $\vec\mu\times\vec{B}$ is perpendicular to the $\hat{z}$ axis.
However, magnetic field {\em gradient} noise couples via:
\begin{equation}
  -m\omega^2 hL = \mu \nabla_x B_z .
\end{equation}
In order to ensure that ambient magnetic fields are not a limiting noise source, it is necessary to limit the magnetic gradient to be $\nabla_x B_z \lesssim \unit[2\times10^{-3}]{fT \, m^{-1} \, Hz^{-1/2}}$.
(Here, we have assumed a magnetic moment of $\mu=\unit[9]{J/T}$.)
Thus, a magnetic shield, placed at a distance of $\approx\unit[10]{m}$ from the torsion magnet, may produce gradients $\lesssim50\times$ above the required noise level.

We consider two strategies to further reduce magnetic noise.
The first is to precisely measure the ambient magnetic field in order to subtract the resulting strain noise.
Warm atomic vapor magnetometers are currently capable of measuring magnetic fields with sensitivities of $\unit[0.5]{fT/Hz^{1/2}}$~\cite{sheng}.
By scaling these sensors to larger volumes, it may be possible to accurately measure residual magnetic gradient noise inside the shielded cavity so that the resulting yaw noise can be canceled by a servo.

The second strategy is to replace each magnetic component with multiple magnets of alternating polarity such that the net magnetic moment of both is zero.
In principle, this can be achieved while still obtaining the magnetic force used to levitate the torsion bar.
With this design, the MAGPI would be sensitive to the second spatial derivative of the magnetic field, potentially reducing the magnetic noise by a factor of $\gtrsim10$ depending on the thickness of the alternating magnets and the precision with which they can be constructed.
The gray~$\bullet$ in Fig.~\ref{fig:noisebudget} show different magnetic noise scenarios.
The upper trace shows the magnetic noise limited only by a magnetic shield.
The lower trace shows the target magnetic noise, reduced through some other means by a factor of $50$.



{\em Additional challenges} are likely.
We discuss a few while acknowledging that our list is certainly incomplete.
First, the motion of the torsion magnet in the external field may incur dissipation through hysteresis.
Second, we expect some degree of noise from the Barkhausen effect.
Third, there is dissipation associated with mechanical stress and strain, especially from resonances created from the attachment of the mirror/magnets to the torsion bar.
Fourth, further study is necessary to understand magneto-mechanical interactions between the torsion bar and the assisting magnets.
Fifth, DeSalvo argues that dislocation self-organized criticality (SOC) noise is a limiting noise source for torsion pendulums~\cite{riccardo}.
However, it might be mitigated through the use of glassy metals.

Sixth, it is not clear if it is feasible to construct a sufficiently stiff torsion bar without significantly increasing the mass of the suspension, which could jeapordize the entire concept.
Seventh, we have made no effort to characterize noise from eddy current damping in the many other (typically) conductive structures that are used in the construction of interferometers including vacuum chambers, support structures, etc.)
Eight, we have assumed perfect magnets; imperfections could induce non-linearities via imperfect magnetic fields.
This, in turn, could lead to problems like the upconversion of out-of-band noise.
Additional work is required to determine the relative importance of these noise sources.
Barkhausen noise is of particular concern given experience from Initial LIGO~\cite{anamaria,barkhausen}.

{\em Conclusions.}
We have carried out a theoretical case study of one possible suspension system for use in a terrestrial millihertz detector.
Inspired by Fig.~7 of~\cite{mango}, we sketched a design for a magnetically assisted pendulum called MAGPI (for Magnetically Assisted Gravitational-wave Pendulum Intorsion).
Significant experimental challenges must be overcome to realize a working MAGPI.
While our proposal (and TOBA~\cite{toba}) utilize the natural resonant frequency of a torsion pendulum, other designs seek to achieve low resonant frequency suspensions using magnetic levitation controlled via active feedback~\cite{varvella,drever}.
Because of the large magnetic fields required for magnetic levitation, eddy current dissipation is an important design challenge for any scheme using magnets.
In our estimation, the challenges posed by magnetic fields are potentially insurmountable.
If it is possible to design a suspension system using a similar principle, but without magnets, the remaining challenges would still be daunting, but perhaps surmountable.

\begin{figure}[hbtp!]
  \includegraphics[width=2.0in]{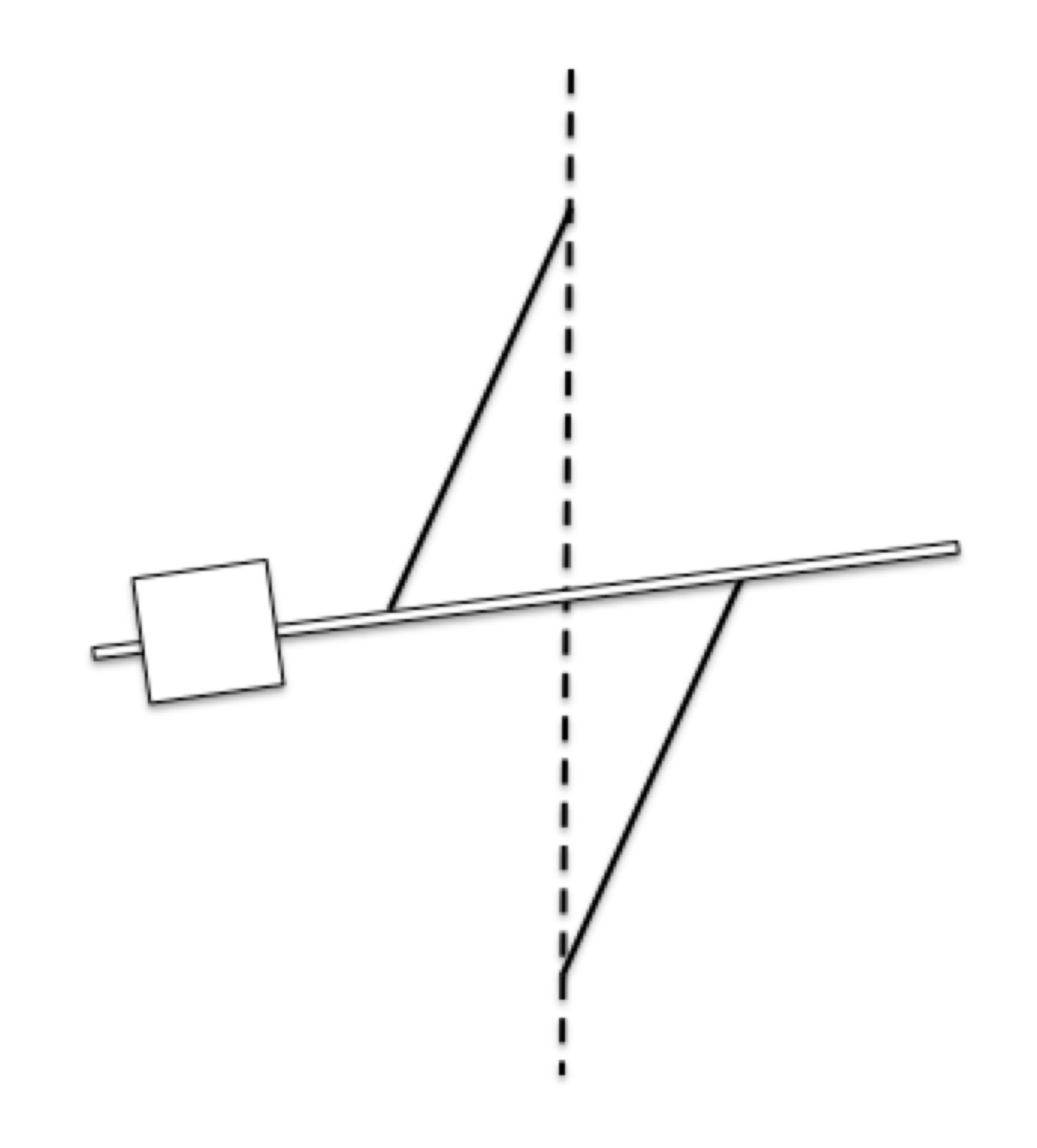}
  \caption{
    Schematic diagram of a Z\"ollner pendulum.
    Like the MAGPI, the low resonant frequency is provided by torsion wire.
    Instead of using magnets to support the asymmetric bar, the Z\"ollner pendulum employs a second wire.
  }
  \label{fig:zollner}
\end{figure}

A promising alternative to MAGPI is a Z\"ollner pendulum, which uses two wires (one above and one below) to suspend a horizontal torsion bar without magnets~\cite{zollner}; see Fig.~\ref{fig:zollner}.
Since the transfer function calculations presented above should also apply to a Z\"ollner pendulum, it would seem that the Z\"ollner design may achieve similar performance to a MAGPI, but without magnetic noise.
We speculate that angular alignment noise is likely still one of many challenging noise sources.
We leave this as future work.
In order to determine the viability of various designs, the next step should be to develop one or more small-scale prototypes (be they MAGPI, Z\"ollner or something else) in order to measure transfer functions and cross-couplings.

We thank Riccardo DeSalvo and Vuk Mandic for discussions, which got us thinking about the problem of low-frequency suspension; Todd Wagner, Vladimir Dergachev, Bram Slagmolen, Rana Adhikari, and John Winterflood for advice; and Lawrie Hanson for the MAGPI acronym.
ET is supported through ARC FT150100281 and CE170100004.
YL is supported through ARC DP1410102578, FT110100384, and CE170100004.

\begin{appendix}
\section{Estimation of eddy current noise}
Here, we derive Eq.~\ref{eq:eddy}, which provides an order-of-magnitude estimate for eddy current damping noise.
Following~\cite{levin}, we use the the fluctuation-dissipation theorem to relate damping associated with yaw motion to eddy-current noise in the yaw degree of freedom; see Fig.~\ref{fig:magpi}.
To begin, we imagine an oscillating force acting directly on the test mass $F(t)=F_0 \cos(\omega t)$ and ask how the resulting motion dissipates heat in the bar magnet.
The oscillating force causes the bar magnet to move with $\hat{x}$ velocity $\tilde{v}_x(\omega) = F_0 / m \omega$.
The tilde denotes that we have moved to the frequency domain.
Here, $m$ is the mass of the test mass since the torsion bar mass is assumed to be comparatively small.

Next, we assume that the field lines move adiabatically so that the induced electric field amplitude in the bar magnet is given by
\begin{equation}
  E^\text{ind} = c \tilde{v}_x B = c \frac{F_0 B}{m \omega} ,
\end{equation}
where $B$ is the magnetic field amplitude from the assisting magnets in the location of the bar magnet.
For the sake of simplicity, we work with single value of $B$, which represents the average field amplitude.
Since we are interested in an order-of-magnitude estimate, we ignore ${\cal O}(1)$ geometric factors, which we absborb into the constant $c$.

The induced electric field generates a current
\begin{equation}
  j = \sigma E^\text{ind} ,
\end{equation}
where $\sigma$ is the conductivity of the bar magnet.
The power dissipated as heat per unit volume is given by
\begin{equation}
  \frac{dP}{dV} = E^\text{ind} j = \frac{c \sigma F_0^2 B^2}{m^2 \omega^2} .
\end{equation}
By assumption, the dissipattion is uniform throughout the bar magnet and so the total dissipated power is
\begin{equation}
  P = c V \left(\frac{\sigma F_0^2 B^2}{m^2 \omega^2}\right) = 
  c \pi R^2 z \left(\frac{\sigma F_0^2 B^2}{m^2 \omega^2}\right) .
\end{equation}
Here, $V=\pi R^2 z$ is the volume of the cylindrically-shaped bar magnet.

Now, we use the fluctuation-dissipation theorem to relate the dissipated power to the power spectral density of length fluctuations in the $\hat{x}$ direction~\cite{levin}.
\begin{equation}
  S_x(\omega) = \frac{8 k_B T}{\omega^2} \frac{P}{F_0^2}  = 
  \frac{8 kB T V \sigma B^2}{m^2 \omega^4} .
\end{equation}
The {\em strain} noise amplitude spectral density is therefore
\begin{equation}
  \sigma_h = S_x^{1/2}/L  \approx c\frac{1}{L} \sqrt{8 k_B T \sigma (\pi R^2 z)} \left(\frac{B}{m \omega^2}\right) .
\end{equation}

\end{appendix}

\bibliography{magpi}


\end{document}